\documentclass[fleqn,twoside]{article}
\usepackage{espcrc2}

\usepackage{graphicx}
\usepackage[figuresright]{rotating}
\usepackage{graphicx,psfrag,amssymb}
\newcommand{\AmS}{{\protect\the\textfont2
  A\kern-.1667em\lower.5ex\hbox{M}\kern-.125emS}}

\title{Direct CP asymmetry of $b\to s\gamma$ and $b\to d\gamma$ 
in models beyond the Standard Model}

\author{A.G. Akeroyd\address{KIAS, Cheongryangri-dong 207-43, 
        Dongdaemun-gu, Seoul 130-012, Republic of Korea},
        S. Recksiegel\address{KEK Theory Group, 
        Tsukuba, Japan 305-0801}}

\begin{document}

\begin{abstract}
We study the direct CP asymmetry of the decays $b\to s\gamma$ and 
$b\to d\gamma$ in the context of two models: i) a supersymmetric (SUSY)
model with unconstrained SUSY phases, and ii) a model with a single 
generation of vector quarks. In both the above models we show that
$b\to d\gamma$ can sizeably influence the combined asymmetry (i.e.\
that of a sample containing both $b\to s\gamma$ and $b\to d\gamma$), 
and in case (ii) may in fact be the dominant contribution.
\vspace{1pc}
\end{abstract}

\maketitle

\section{Introduction}

Theoretical studies of rare decays of $b$ quarks have attracted
increasing attention since the start of the physics programme at the 
$B$ factories at KEK and SLAC. 
Many rare decays will be observed for the first time
over the next few years, and in this talk we summarize our 
work on the decays $b\to d\gamma$ and $b\to s\gamma$ in the context
of two models: i) a supersymmetric (SUSY)
model with unconstrained SUSY phases \cite{Akeroyd:2001cy}, and 
ii) a model with a single generation of vector quarks \cite{Akeroyd:2001gf}.

There is considerable motivation for measuring the 
BR and CP asymmetry (${\cal A}_{CP}$) of the inclusive channel
$B\to X_d\gamma$. In particular we highlight the following:

\begin{itemize}

\item[{(i)}] $b\to d\gamma$ transitions sizeably affect
the measurements of ${\cal A}_{CP}$ for $b\to s\gamma$ \cite{Coan:2001pu}.
Therefore 
knowledge of ${\cal A}_{CP}$ for $b\to d\gamma$ is essential, in order to
compare experimental data with the theoretical prediction 
in a given model. 

\item[{(ii)}] ${\cal A}_{CP}$ for the 
combined signal of $B\to X_s\gamma$ and 
$B\to X_d\gamma$ is expected to be close to zero in the 
Standard Model (SM)
\cite{Soares:1991te,Kagan:1998bh,Hurth:2001yb}, due the
real Wilson coefficients and the unitarity of the CKM matrix. 
Both of these conditions can be relaxed in models beyond the SM.

\end{itemize}

\section{The decays $b\to d\gamma$ and $b\to s\gamma$}

There is much theoretical and experimental motivation to
study the ratio
\begin{equation}
R={BR(B\to X_d\gamma)\over {BR(B\to X_s\gamma)}}
\end{equation}
because it provides a clean handle on the ratio
$|V_{td}/V_{ts}|^2$ \cite{Ali:1998rr}.
In the context of the SM, $R$ is expected to be in the 
range $0.017 < R < 0.074$, corresponding
to BR$(B\to X_d\gamma)$ of order $10^{-5}$. 
$R$ stays confined to this range 
in many popular models beyond the SM. This is because new 
particles such as charginos and charged Higgs bosons in SUSY models
contribute to $b\to s(d)\gamma$ 
with the same CKM factors. Therefore $C_7$ is universal to both
decays and cancels out in the ratio $R$. 
In a model with vector quarks this is not the case, and we 
shall see that $R$ can be suppressed or enhanced 
with respect to the SM.

${\cal A}^{d\gamma(s\gamma)}_{CP}$ is given by:
\begin{equation}
{{\Gamma(\overline B\to X_{d(s)}\gamma)-
\Gamma(B\to X_{\overline d(\overline s)}\gamma)}
\over {\Gamma(\overline B\to X_{d(s)}\gamma)+\Gamma(B\to X_{\overline
d(\overline s)}\gamma)}}={\Delta\Gamma_{d(s)}\over \Gamma^{tot}_{d(s)}}
\label{ACPdef} \end{equation}
In the SM ${\cal A}_{CP}^{d\gamma}$ is expected to lie in the 
range $-5\%\le {\cal A}_{CP}^{d\gamma}\le -28\%$ 
\cite{Ali:1998rr}, where the uncertainty arises from 
varying the Wolfenstein parameters 
$\rho$ and $\eta$ in their allowed ranges. 
Therefore  ${\cal A}_{CP}^{d\gamma}$ is much larger than
${\cal A}_{CP}^{s\gamma}$ ($\le 0.6\%$). 

If $b\to d\gamma$ and $b\to s\gamma$ 
cannot be properly separated, then only $A_{CP}$ of
a combined sample can be 
measured. It has been shown \cite{Soares:1991te,Kagan:1998bh,Hurth:2001yb}
that ${\cal A}_{CP}^{s\gamma}$ and
$A_{CP}^{d\gamma}$ approximately cancel each other in the SM, 
leading to a combined asymmetry close to zero.

A reliable prediction of 
${\cal A}^{d\gamma}_{CP}$ in a given model is necessary 
since it contributes to the measurement of ${\cal A}^{s\gamma}_{CP}$.
The CLEO result \cite{Coan:2001pu} is sensitive to a weighted sum of 
CP asymmetries, given by:
\begin{equation}
{\cal A}^{exp}_{CP}=0.965{\cal A}^{s\gamma}_{CP}+0.02{\cal A}^{d\gamma}_{CP}
\label{CLEOeq} \end{equation}  
The latest measurement stands at $-27\% < {\cal A}^{exp}_{CP} < 10\%$ 
(90\% C.L.) \cite{Coan:2001pu}. 
The small coefficient of ${\cal A}^{d\gamma}_{CP}$ is caused by
the smaller BR$(B\to X_d\gamma)$ (assumed to be $1/20$ that
of BR$(B\to X_s\gamma)$) and inferior detection efficiencies.

If the detection efficiencies for both decays were identical,
this measured quantity would coincide with the weighted sum of the asymmetries
\begin{equation}
{\cal A}_{CP}^{s\gamma+d\gamma} = { {\rm BR}^{s\gamma} {\cal A}_{CP}^{s\gamma}
 +{\rm BR}^{d\gamma} {\cal A}_{CP}^{d\gamma} \over {\rm BR}^{s\gamma} +{\rm BR}^{d\gamma}}\,.
\label{ACPcombined} \end{equation}  

The two terms in eqs.~(\ref{CLEOeq},\ref{ACPcombined}) can
be of equal or of opposite sign, i.e.\ they can contribute 
constructively or destructively to the combined asymmetry.
The non--negligible contribution of
$b\to d\gamma$ to this combined asymmetry 
should be verifiable at proposed future high luminosity  
runs of $B$ factories.

\section{Results}
We now show numerical results for the two models considered. 
\subsection{Effective SUSY model}

In Fig.~1 we plot ${\cal A}^{d\gamma}_{CP}$ against 
$m_{\tilde t_1}$, which clearly shows that a light $\tilde t_1$ may 
drive ${\cal A}_{CP}^{d\gamma}$ positive, reaching maximal 
values close to $+40\%$. For $\tilde t_1$ heavier than 250 GeV
the ${\cal A}_{CP}^{d\gamma}$ lies within the SM range, which
is indicated by the two horizontal lines.
In Fig.~2 we plot ${\cal A}_{CP}^{d\gamma}$ against 
${\cal A}_{CP}^{s\gamma}$.
One can see that both ${\cal A}_{CP}^{s\gamma}$ and
${\cal A}_{CP}^{d\gamma}$ may have either sign, resulting
in constructive or destructive interference in eq.~(\ref{CLEOeq}).

In Fig.~3 we plot the ${\cal A}_{CP}^{exp}$ 
(defined in eq.~(\ref{CLEOeq})) against ${\cal A}_{CP}^{s\gamma}$. 
If the contribution from ${\cal A}_{CP}^{d\gamma}$ were ignored
in eq.~(\ref{CLEOeq}), then Fig.~3 would be a straight 
line through the origin. The ${\cal A}_{CP}^{d\gamma}$ contribution 
broadens the line to a thin band of width $\approx 1\%$, an effect
which should be detectable at proposed higher luminosity runs of the
$B$ factories.

\subsection{Vector quark model}

In Fig.~4 we plot ${\cal A}^{d\gamma}_{CP}$
against ${\cal A}^{s\gamma}_{CP}$.
It can be seen that ${\cal A}^{s\gamma}_{CP}$
does not substantially differ from its SM value, 
while ${\cal A}^{ d\gamma}_{CP}$ can vary over a much larger
range.  The correlation between ${\cal A}^{d\gamma}_{CP}$ and
BR$^{d\gamma}$ is studied in detail in Fig.~5,
where it can be seen that $|{\cal A}^{ d\gamma}_{CP}| >45 \%$
occurs only for BR$^{d\gamma} < 10^{-6}$. Branching ratios
of this magnitude would require $\gg 10^8$ $b\bar b$ pairs
to be detected which is beyond the discovery potential of
current $B$ factories.

In Fig.~6  we plot the combined CP asymmetry as defined
in eqn.(\ref{ACPcombined}) against the argument of $V_{Ud}^*V_{Ub}$.
Note that in our analysis BR$^{s\gamma}$ and 
${\cal A}^{s\gamma}_{CP}$ are close to their SM values.
The huge variations in ${\cal A}^{s\gamma+d\gamma}_{CP}$
stem from the variation in BR$^{d\gamma}$. In wide ranges
of our parameter space, $b\to d\gamma$ actually dominates
the combined asymmetry ! Any large signal observed in
${\cal A}^{s\gamma+d\gamma}_{CP}$ is a sign of physics
beyond the SM, but although  BR$^{s\gamma+d\gamma}$ is
strongly dominated by $b\to s \gamma$, a non--SM value
for ${\cal A}^{s\gamma+d\gamma}_{CP}$ can stem from both
$b\to s \gamma$ and $b\to d \gamma$.

\newpage

\begin{figure}
\begin{center}
\psfrag{XXX}{$m_{\tilde t_1}$}  \psfrag{YYY}
 {${\cal A}^{d\gamma}_{CP}$}
\includegraphics[width=7.5cm]{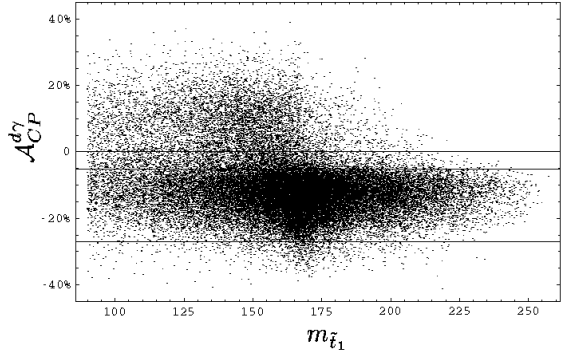}
\end{center}
\vspace*{-1.2cm}
\caption{${\cal A}^{d\gamma}_{CP}$ against $m_{\tilde t_1}$}
\label{mst1}
\end{figure}

\begin{figure}
\begin{center}
\psfrag{XXX}{${\cal A}^{s\gamma}_{CP}$}  \psfrag{YYY}
 {${\cal A}^{d\gamma}_{CP}$}
\includegraphics[width=7.5cm]{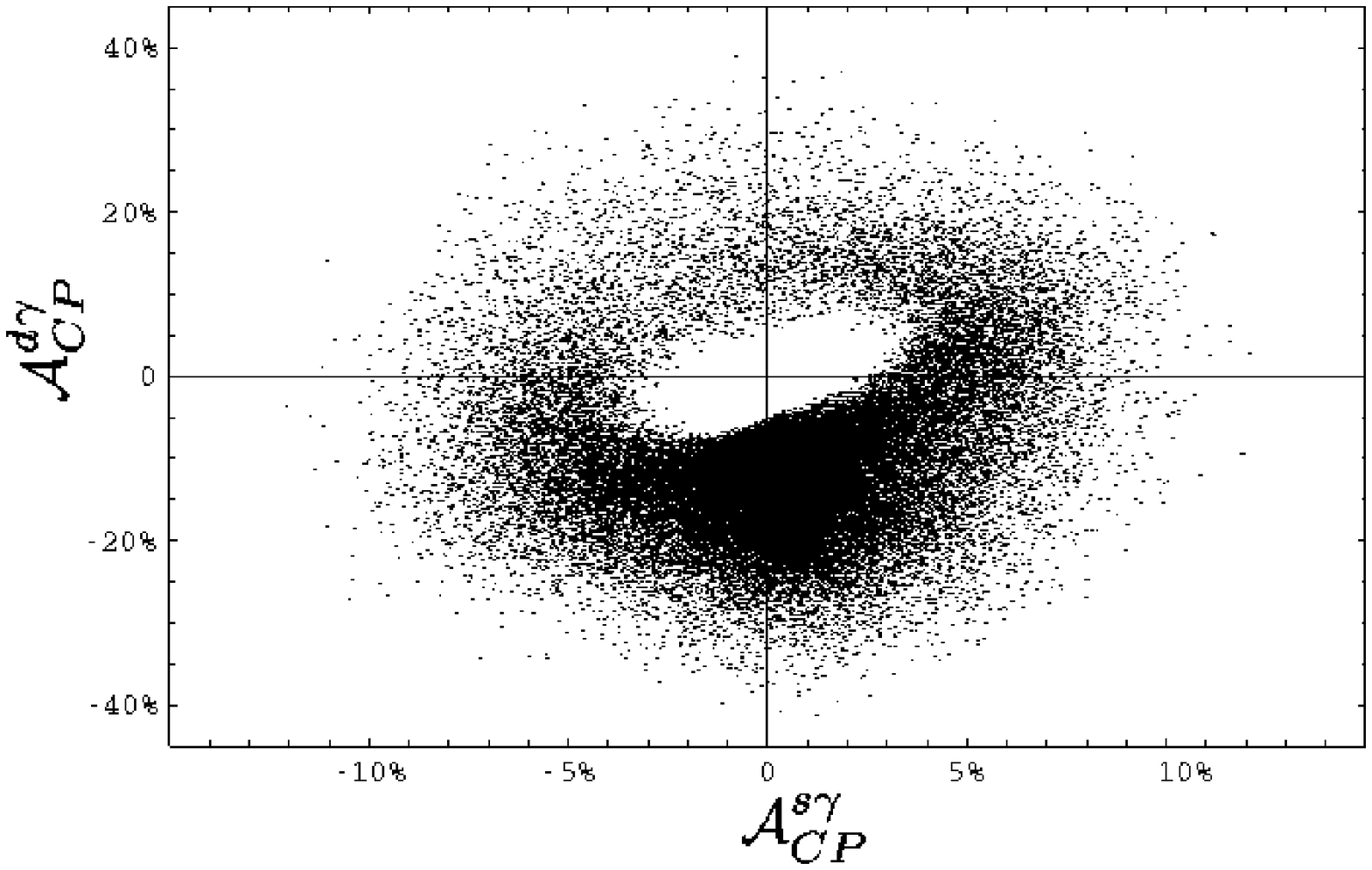}
\end{center}
\vspace*{-1.2cm}
\caption{${\cal A}^{d\gamma}_{CP}$ against ${\cal A}^{s\gamma}_{CP}$}
\label{doughnut}
\end{figure}
\begin{figure}
\begin{center}
\psfrag{XXX}{${\cal A}^{s\gamma}_{CP}$}  \psfrag{YYY}
 {${\cal A}^{exp}_{CP}$}
\includegraphics[width=7.5cm]{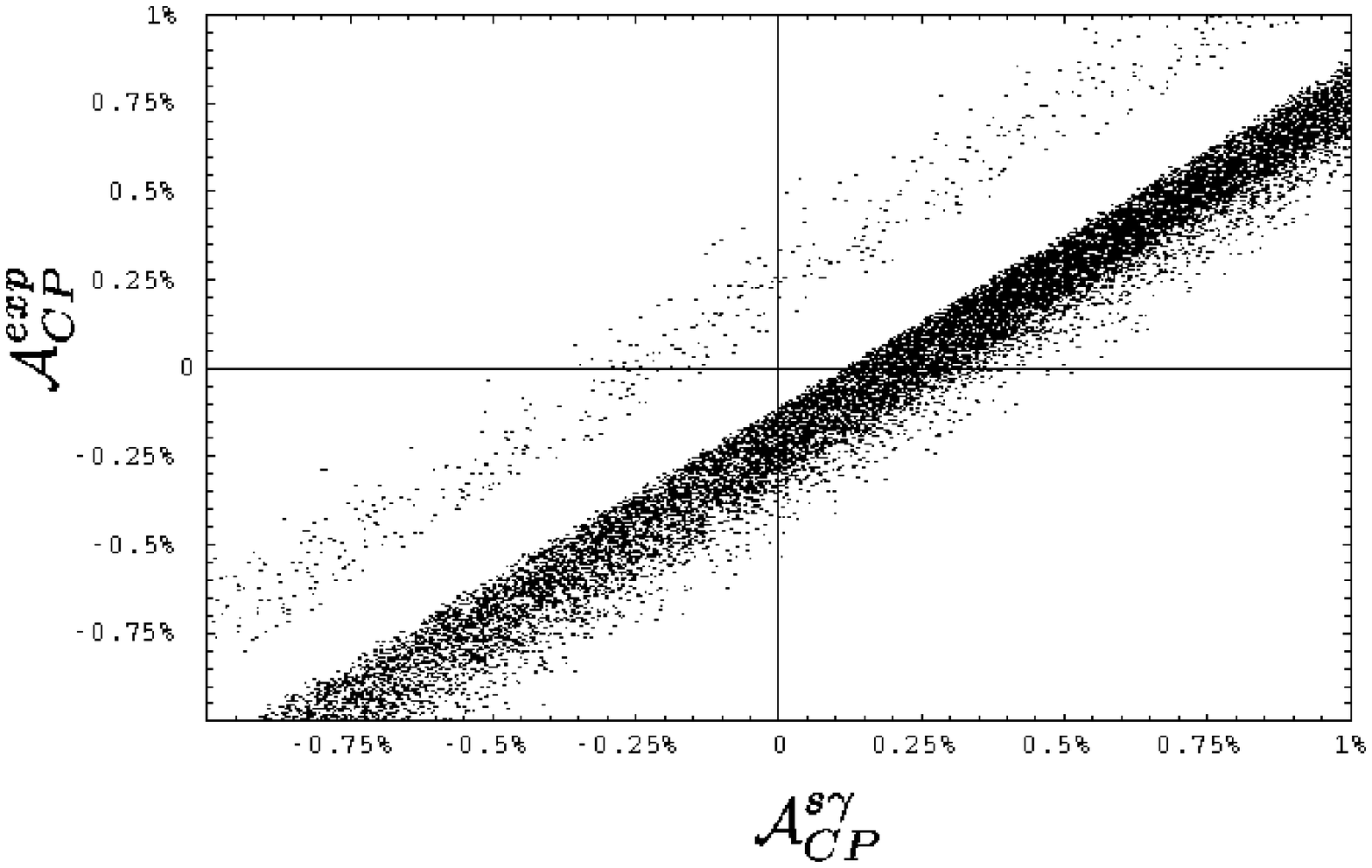}
\end{center}
\vspace*{-1.2cm}
\caption{CLEO ${\cal A}^{exp}_{CP}$ against 
 ${\cal A}^{s\gamma}_{CP}$}
% ${\cal A}^{s\gamma}_{CP}$}
\label{ACPCLEO}
\end{figure}

\begin{figure}
\begin{center}
\psfrag{XXX}{${\cal A}^{\rm s\gamma}_{CP}$}  \psfrag{YYY}
 {${\cal A}^{\rm d\gamma}_{CP}$}
\includegraphics[width=7.7cm]{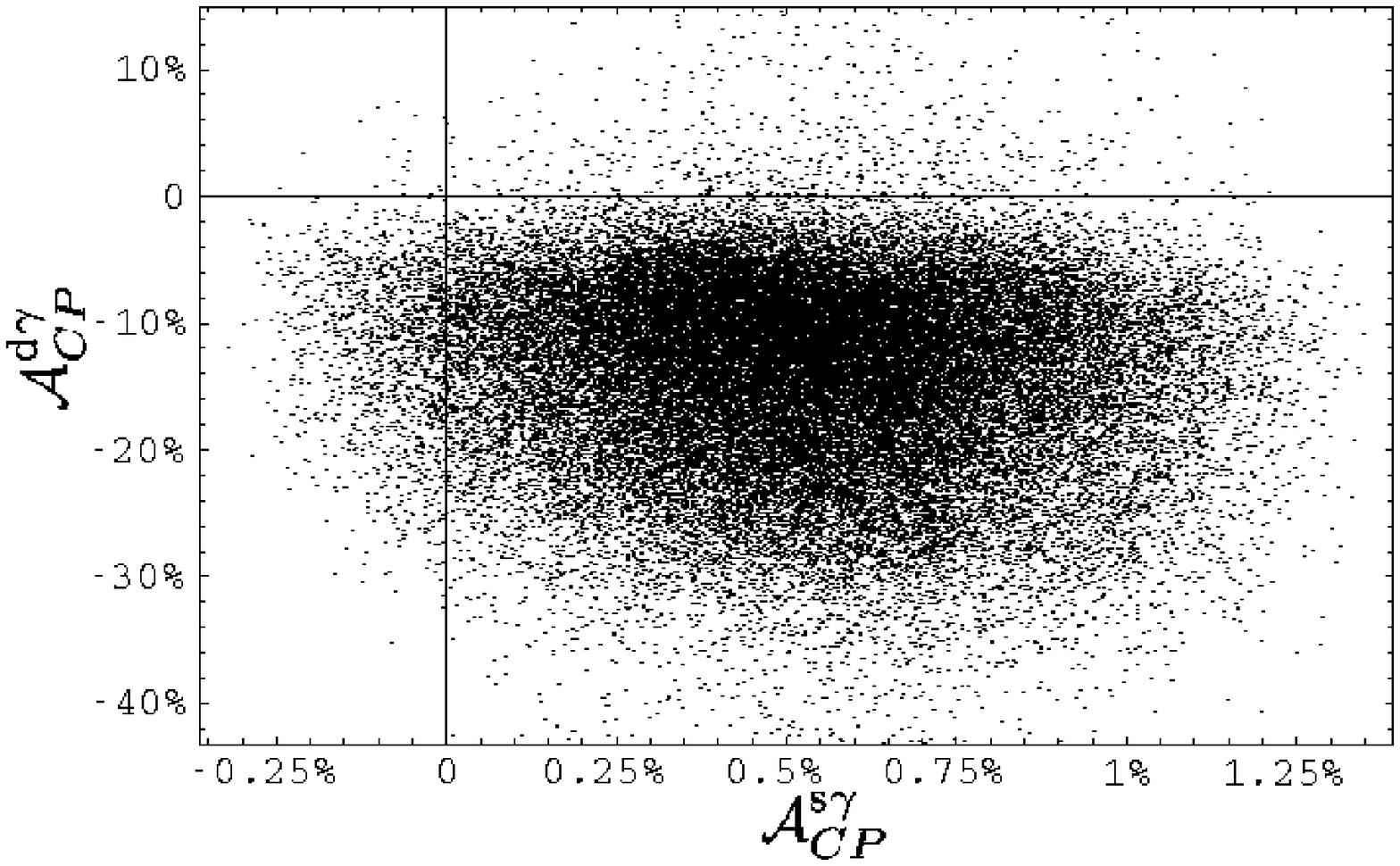}
\end{center}
\vspace*{-1.2cm}
\caption{${\cal A}^{d\gamma}_{CP}$ against ${\cal A}^{s\gamma}_{CP}$}
\label{scatter}
\end{figure}

\begin{figure}
\begin{center}
\psfrag{XXX}{BR$(b\to d\gamma)\,[10^{-5}]$}  \psfrag{YYY}
 {${\cal A}^{d\gamma}_{CP}$} \psfrag{ZZZ} {$|V_{Ud}^*V_{Ub}|$} 
\includegraphics[width=7.5cm]{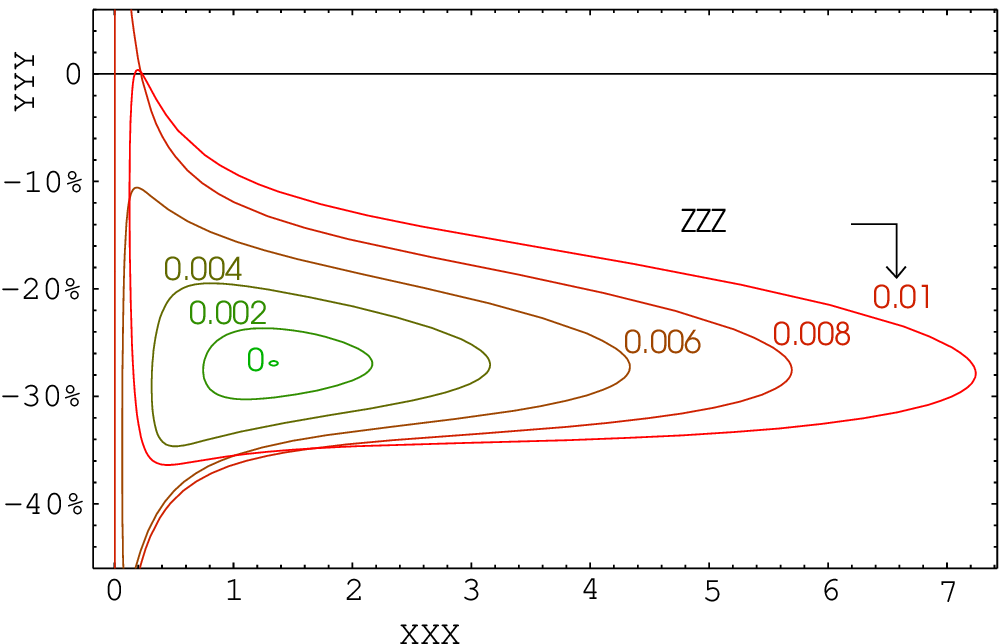}
\end{center}
\vspace*{-1.2cm}
\caption{${\cal A}^{d\gamma}_{CP}$ against
BR($b\to d\gamma$)} 
\label{contours}
\end{figure}

\begin{figure}
\begin{center}
\psfrag{XXX}{Arg $V_{Ud}^*V_{Ub}\,\,[\pi]$}  \psfrag{YYY}
 {${\cal A}^{s\gamma+d\gamma}_{CP}$} \psfrag{ZZZ} {$|V_{Ud}^*V_{Ub}|$}
\includegraphics[width=7.5cm]{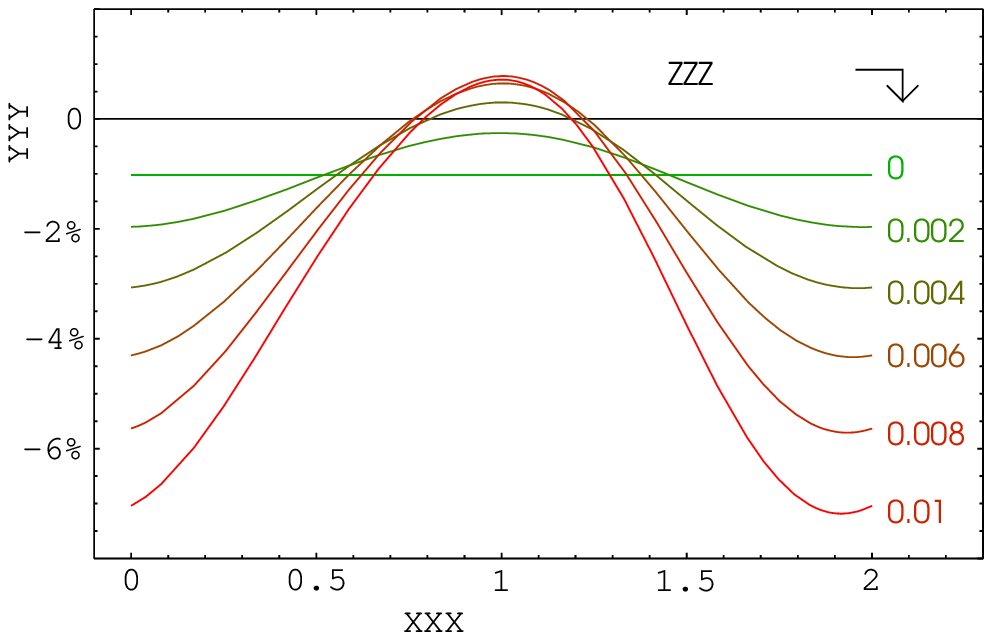}
\end{center}
\vspace*{-1.2cm}
\caption{${\cal A}^{s\gamma+d\gamma}_{CP}$ against Arg $V_{Ud}^*V_{Ub}$}
\label{hill}
\end{figure}

\end{document}